# Longitudinal phase-space coating of beam in a storage ring

C. M. Bhat

Fermi National Accelerator Laboratory, P.O.Box 500, Batavia, IL 60510, USA

Phone : 630-840-4821

Fax : 630-840-8461

Email : cbhat@fnal.gov

**Abstract**

In this Letter, I report on a novel scheme for beam stacking without any beam emittance dilution using a barrier rf system in synchrotrons. The general principle of the scheme called longitudinal phase-space coating, validation of the concept via multi-particle beam dynamics simulations applied to the Fermilab Recycler, and its experimental demonstration are presented. In addition, it has been shown and illustrated that the rf gymnastics involved in this scheme can be used in measuring the incoherent synchrotron tune spectrum of the beam in barrier buckets and in producing a clean hollow beam in longitudinal phase space. The method of beam stacking in synchrotrons presented here is the first of its kind.

## 1. Introduction

With the recent surge in world-wide interest in the *beam intensity frontier*, high intensity beam stacking in synchrotrons without any emittance dilution has become very important. Several decades of research have resulted in many novel methods of beam accumulation in storage rings [1-7]. However, during the Tevatron collider operations at the Fermilab between 2000 to 2011 none of the previously developed beam stacking methods could be used in the Recycler Ring [8], a permanent magnet 8 GeV storage synchrotron, which exclusively used *barrier radio-frequency* (rf) systems [9, 10] in all of its beam manipulations. For this reason and in order to support luminosity upgrades [11] new methods needed to be developed for beam operation[12, 13].

The barrier rf systems were introduced to modern accelerator technology by Griffin et. al., [9] in 1983 and the flexibility they provided in the handling the beam with very complex bunching formats, like that demonstrated in longitudinal momentum mining [14], has opened up new prospects in beam dynamics not possible with conventional harmonic rf systems. Over the past decade, substantial experimental and theoretical progress have taken place in the field of barrier rf systems because of their important role in the Fermilab Recycler Ring [12-16], induction accelerator at KEK [17], R&D effort at CERN and BNL [18] and the proposed future NESR facility at GSI [19]. Despite these developments, not enough attention was given on improving beam stacking using barrier rfs. In the case of the Recycler, the antiproton beam was accumulated by repeated beam transfers from the Pbar source and cooled throughout the beam accumulation. It was imperative to keep the emittance of the cooled beam intact during the rf manipulations needed during beam stacking. All of the previously adopted beam accumulation techniques [12, 13] encountered difficulty in preserving longitudinal emittance.

In this Letter, we demonstrate a novel beam accumulation scheme called "longitudinal phase-space coating" (LPSC) [20]. The phase space density of the initial beam is held unchanged and the emittance dilution for the newly arrived beam kept minimal. The physics concepts, an experimental demonstration and some spin-off applications of the scheme are presented.

## 2. Formalism of LPSC

The Hamiltonian for the synchrotron motion of a particle in a barrier bucket made of an arbitrary rf wave form $V(t)$ can be written as [15],

$$H(\Delta E, \tau) = -\frac{\eta}{2\beta^2 E_0} \Delta E^2 - \frac{e}{T_0}\int_0^\tau V(t)dt \qquad (1)$$

where the quantities $E_0, \Delta E, e, \eta, T_0$ and $\beta$ represent synchronous energy, energy offset from $E_0$, electronic charge, phase slip factor, revolution period and the relativistic velocity of the particle, respectively. The time difference between the arrival of the particle and that of a synchronous particle at the center of the rf bucket is denoted by $-\tau$. The second term in Eqs. (1) represents the potential energy of the particle. In the absence of non-linear forces like *intra-beam scattering* and/or *synchro-betatron coupling*, a particle will continue to follow the contour of a constant Hamiltonian in an rf bucket performing synchrotron oscillations. The maximum energy offset $\Delta E_{Max}$ of a particle during its synchrotron motion is related to its penetration depth $T_{Max}$ into the barrier, and is given by,

$$\Delta E_{Max} = \sqrt{\frac{2e\beta^2 E_0}{|\eta|T_0}\left|\int_{T_2/2}^{T_2/2+T_{Max}} V(t)dt\right|} \qquad (2),$$

assuming rf pulses are anti-symmetric with respect to the center of the bucket. $T_2$ denotes the gap between positive and negative pulses that forms a barrier bucket. In the case of rectangular barrier pulses one can replace the integral in this equation by $V_0 T_{Max}$ where $V_0$ represents maximum of rf pulse height. Further, any rectangular barrier bucket can be treated as a combination of multiple buckets, one inside the other, like a *matryoshka* *doll*, so that one of the inner buckets confines all of the particles below $\Delta E_{Max}$. The principal goal of the new stacking scheme is to isolate all the particles below a certain maximum energy spread using an inner barrier bucket (*mini-barrier* bucket). The maximum potential energy of the particles in the mini-bucket is set at the same level as the minimum potential energy of the newly injected beam and coat the new beam on top of the isolated beam. The coating takes place in $(\Delta E, \tau)$–space. The particles in the mini-bucket will be left undisturbed throughout the stacking.

Figure 1 provides an insight into the coating mechanism. A schematic view of the rf wave-forms with the beam phase space boundary and the beam particles in the corresponding potential well for a storage ring operating below transition energy are shown. Before the transfer of a new beam, a mini-bucket made of two barrier pulses "3" and "4" is opened adiabatically to isolate the initial beam, as shown in Fig. 1(b). The mini-bucket isolates particles in a phase space area $\varepsilon_m = 2(T_2 - 2T_m)\Delta E_m + 4T_0 |\eta| \Delta E_m^3 / [3\beta^2 E_0 eV_m]$. The quantities $T_m$, $\Delta E_m$ and $V_m$ represent pulse width, bucket height and pulse height for the mini-bucket, respectively. Subsequently, a new beam is injected in a separate barrier bucket made of rf pulses "5" and "6". For simplicity, the parameters of barrier pulses "5" and "6" are chosen same as those of "1" and "2", respectively. The rest of the beam stacking involves adiabatic and simultaneous removal of the barrier pulses "2" and "5". Thus, the injected beam particles start slipping along the newly formed contours of constant Hamiltonian around the mini-bucket as depicted in Fig. 1(c). Eventually, the rf pulse "6" is moved to the location of "2" adiabatically to complete the coating process as shown in Fig. 1(d).

## 3. Experimental demonstration

We have demonstrated the above beam stacking scheme in the Recycler which operated below the transition energy of 20.27 GeV with $T_0 = 11.12$ μsec. The Recycler barrier rf system was capable of providing pulses of any shape with amplitudes up to ~2 kV [10] and had a versatile LLRF [16]. The 2D- particle tracking simulation code ESME [21] was employed to validate the scheme and to establish the rf manipulation steps. Subsequently, the beam experiments were carried out as proof of principle demonstration.

Figure 2 shows simulated beam particle distributions in the longitudinal phase space along with the barrier rf pulses. An initial beam of ~125 eV s was populated in a barrier bucket of total area 250 eV s with pulse width $T_1 = 0.91\,\mu\text{sec}$, $T_2 = 5.89\,\mu\text{sec}$ and $V_0 = 1.93$ kV at clock time =0 sec. A new beam of about 7 eV s was injected after confining 108 eV s of the initial beam. The parameters of the mini-barrier bucket were $T_m = 0.25$ μsec, $(T_2 - 2T_m) = 5.4$ μsec, $V_m = 1.93$ kV and $\Delta E_m = 9.41$ MeV as shown in Fig. 2(b). There are an infinite number of ways of selecting the parameters for a mini-bucket of phase space area of 108 eV s. The principal idea is that the mini-bucket must isolate the entire initial beam or some part of it. The rest of the beam coating was performed as depicted in Fig. 1. In the simulations, the width of the barrier pulses separating the initial and the new beam were reduced symmetrically to the "unstable point" (see Fig. 2(b)), keeping the newly arrived beam undisturbed. To complete the coating, the right-most barrier pulse was moved adjacent to the mini-barrier bucket as in Fig. 2(d). The simulation showed that the final emittance of the beam was about 132 eV s with a negligible emittance dilution. Repeating the steps shown in Figs. 2(b) to 2(d) one can perform multiple coatings. The simulations clearly showed that the emittance preservation in this scheme depends on the iso-adiabaticity of the rf manipulation. An initial distribution with Gaussian-like tails in both independent coordinates, *viz.,* $\Delta E$ and time, is not very favorable for coating, though the beam in the mini-bucket is undisturbed. However, if the tail is non-Gaussian like that exhibited by the Recycler beam, the coating can be performed with no emittance growth.

Now I discuss the detailed beam measurements confirming the simulation results. The beam tests were carried out using proton as well as antiproton beams using a range of initial beam intensities and following the rf manipulation steps explained in the simulation example described in Fig. 2. Starting with an initial beam of $2.56 \times 10^{12}$ antiprotons and a longitudinal emittance of 70±7 eV s (95%), about 66 eV s of dense region of the beam was isolated using a mini-barrier bucket with $V_m = 0.72$ kV and $\Delta E_m = 5.8$ MeV (which contained ~93% of the initial beam). Subsequently, three coatings were applied consecutively. The longitudinal emittance and beam intensity for these three coatings were 7±1, 8±1 and 7±1 eV s with $14 \times 10^{10}$, $9 \times 10^{10}$ and $5 \times 10^{10}$ antiprotons, respectively. No particle losses were observed during and after coatings. The measured wall current monitor and the Schottky data (red traces) for the beam after completion of coating are shown in Figs. 3(a) and 3(b), respectively. The longitudinal emittance of the final beam was measured to be 100±10 eV s (95%). We followed Monte Carlo methods [22] to measure the longitudinal emittances for the initial as well as final beam distributions. The 95% contour in the final phase space distribution is shown by the blue curve in Fig. 3(c). By adding the errors in quadrature, we find that the observed overall emittance dilution is within the measurement errors of about 10% of the experiment. The comparison between predicted and the measured data is quite satisfactory.

**4. Incoherent spectrum of the beam in barrier buckets**

One of the essential steps in the quantitative understanding of single particle longitudinal beam dynamics in an rf bucket involves an extensive measurement of the synchrotron tune spectrum. Such a measurement has been carried out for particles in a sinusoidal rf bucket at the Indiana University Cyclotron Facility cooler [23]. The LPSC method of beam stacking provides an excellent way to measure the synchrotron tune of the beam in a barrier bucket of any shape. A barrier bucket with $V_0 = 1.84$ kV and with all other parameters similar to the one shown in Fig. 2(a) is chosen for this demonstration. Measured waveform of the bucket is shown in Fig. 4(a). With no beam in the Recycler, a mini-barrier bucket with bucket length of $T_2$ was grown inside the main bucket. Then about $15 \times 10^{10}$ protons were coated on to the empty mini-bucket. The incoherent synchrotron frequency $f_s$ of the beam particles on the outermost separatrix of a mini-rectangular barrier bucket is

$$f_s^{-1} = \frac{2(T_2 - 2T_m)\beta^2 E_0}{|\eta \Delta E_m|} + \frac{4T_0 |\Delta E_m|}{eV_m} \qquad (3).$$

This created a long bunch with no beam particles in the central region of the longitudinal phase space. The separatrix of the mini-bucket acted as the inner boundary between the empty region and the coating. Typical Schottky data from such a beam will have two *horn*-like humps as shown in Fig. 4(b). The Fourier analysis of the wall current monitor signal (using an Agilent 89441A 2.65 GHz Vector Signal Analyzer) gave the synchrotron frequency of the particles on the separatrix. Figure 4 shows measured $|\Delta E_m|$ versus $f_s$ along with the analytical predictions, assuming an ideal rectangular barrier bucket. For $T_2 > 4T_1$ the synchrotron tune is expected to be a monotonic function of $\Delta E$ [15]; the measurement depicts similar behavior. The level of discrepancy between the measurements and the predictions of synchrotron frequency tune can be attributed to the systematic errors in the measurerements of $V_0$ and the difference between exact shapes of the rf pulses used in the experiment and in the calculations.

## 5. Hollow-bunch in barrier bucket

During each of the synchrotron tune measurements mentioned above, a hollow-bunch is created. However, we found that a clean hollow-bunch can be maintained even if the mini-barrier bucket was turned off after its formation as shown in Fig. 5. Figure 5(c) shows a reconstructed hollow-bunch. We see a good agreement between the predictions and the measurement data for the line-charge and energy distributions as shown in Figs. 5(a) and (b), respectively.

## 6. Remarks on high intensity beam stacking

The simulations presented in Figs. 2 were carried out using single-particle beam dynamics without including beam space-charge or wake field effects. Generally these simulations are sufficient to describe the proof of principle of the technique. However, to better understand the high intensity effects one must include the collective effects in the simulations. The simulation results presented in Fig. 3, on the other hand, include i) space charge impedance, ii) inductive impedance from the entire ring which takes into account the reaction of the beam environment on the beam distribution and iii) longitudinal coupling impedances arising from the rf cavity (a shunt impedance of 200 Ω from all four Recycler-barrier rf cavities with Q=1) after full compensation. The total impedance $Z(\omega)$ seen by a Fourier component of the beam current at a frequency $\omega/2\pi$ is modeled as the sum of all three terms. Such analyses when carried out without compensating for the coupling impedance gave rise to a potential well distortion which led to an asymmetric line-charge distribution in long bunches that was in agreement with our early studies [24]. To eliminate this problem, a FPGA-based LLRF adaptive feed-back system was implemented [25] and was used successfully in operation for antiproton intensities in excess of $500 \times 10^{10}$. Further simulations with full compensation for the cavity impedance showed that in the case of the Recycler the LPSC scheme works well without any detrimental effects for intensities in excess of twice the intensity used operationally.

## 7. Summary

In conclusion, this letter proposes and validates a novel beam stacking method for a storage ring that uses barrier rf buckets. It is important to note that the beam stacking described here can use any of the barrier rf waveforms illustrated in ref. 15. We have demonstrated the ability of this technique for beam stacking in the Recycler with an emittance dilution of ~10% per three beam transfers and successfully commissioned the technique in the collider operation at Fermilab. (For comparison, without the LPSC the observed emittance dilution for three consecutive transfers in the Recycler used to be as high as ~45%.)   Also, if the beam is cooled during stacking with the LPSC scheme by separating the core of the beam from the tail of the longitudinal phase space distribution similar to one shown in Fig. 1(c), one may be able to increase the overall cooling rate significantly [26].  Even if beam cooling is not the part of the facility, the LPSC helps to keep the majority of the beam undisturbed as demonstrated here.

The rf gymnastics involved in the longitudinal phase space coating scheme can be utilized in creating an ideal hollow beam in longitudinal phase space. We have demonstrated its use in measuring the incoherent synchrotron tune spectrum of the beam in a barrier bucket. Agreement between the measured and predicted synchrotron tune spectrum is quite satisfactory. It is foreseen that the hollow beam explained here may pave a path to study beam distribution functions as well as the beam dynamics issues of long bunches created using symmetric/asymmetric barrier rf buckets.

As a final note, the activation of environment surrounding the hadron accelerator due to beam loss is generally quite serious, so the LPSC could be an attractive not only for high intensity beam stacking but also for suppression of the activation. Further, we expect that the applications of the LPSC technique described here or some variation of this might benefit planned heavy ion storage rings and other low energy circular storage rings as well.  Hence, it is highly desirable to carry out in-depth studies along these lines.

**Acknowledgement**

The author would like to thank D. Neuffer, D. Wildman P. Bhat and Jim MacLachlan for many useful discussions. Special thanks are due to Shreyas Bhat and Nita Bhat for careful editing of the manuscript and the Fermilab Accelerator Division MCR crew for their help during the beam experiments. This work is supported by Fermi Research Alliance, LLC under Contract No. DE-AC02-07CH11359 with the U.S. Department of Energy.

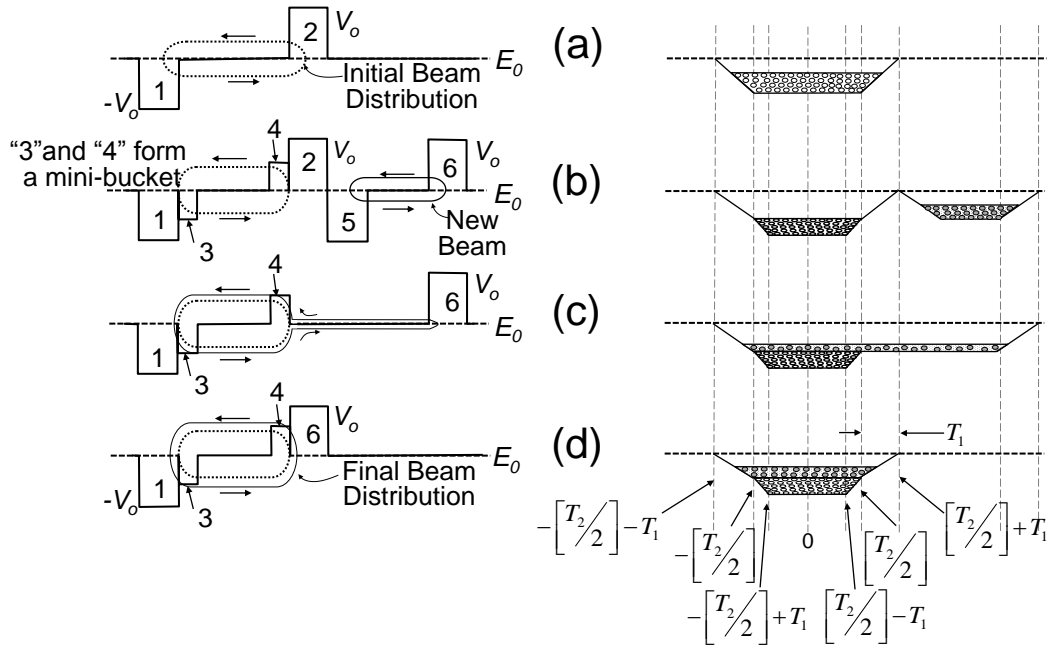

Fig. 1. Schematic view of LPSC. (a) Initial beam, (b) initial beam in a mini-barrier bucket and the new injection, (c) a stage of coating after removal of barrier pulses "2" and "5", and (d) after coating. The voltage wave forms (solid lines) and direction of the synchrotron motion are also shown on the left half of the figure.

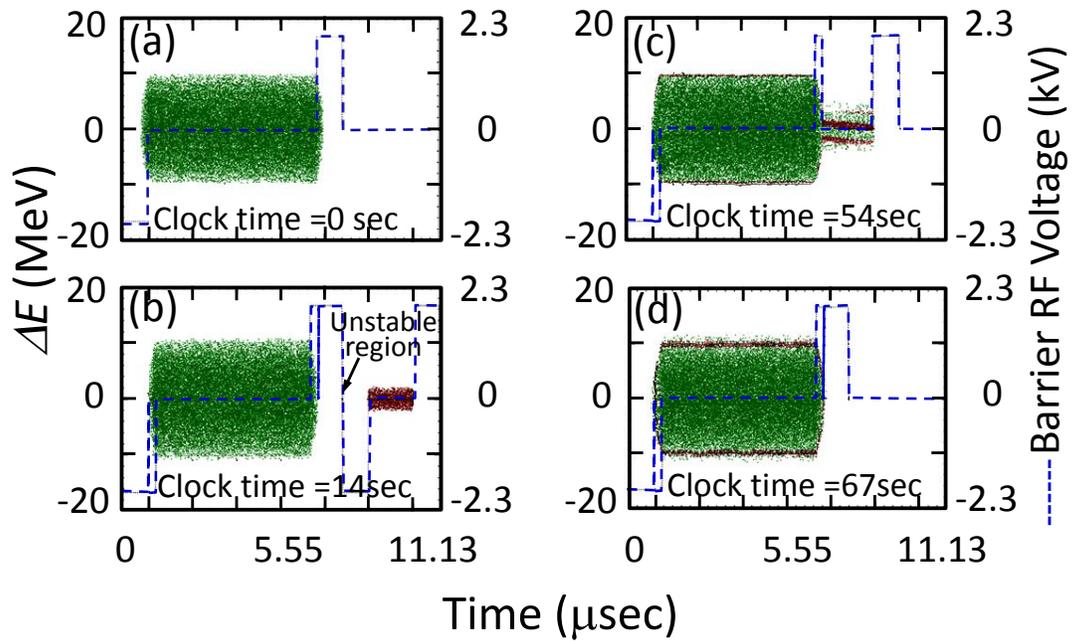

Fig. 2. Simulations of longitudinal phase space coating in the Recycler. The beam distributions and rf wave forms are displayed. The stages of coating are similar to those presented in Fig. 1. The time taken for rf manipulations are also shown.

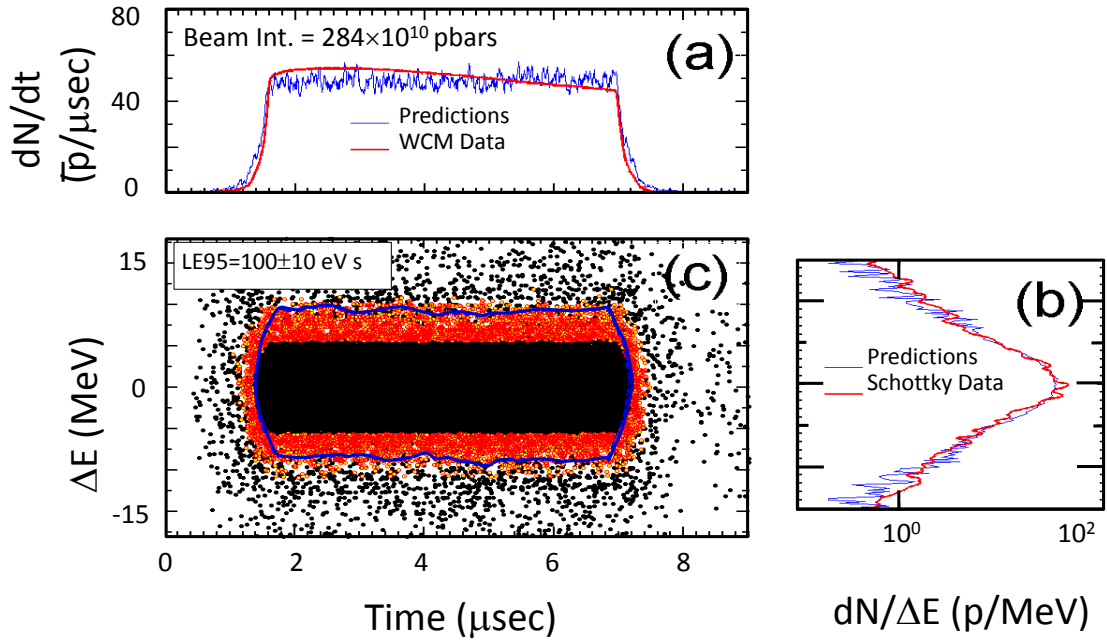

Fig. 3. An overlay of experimental (red curve) and simulation (blue curve) results corresponding to the (a) wall current monitor and (b) Schottky detector data after three coats on an initial beam of $2.56 \times 10^{12}$ antiprotons. (c) Reconstructed longitudinal phase-space distribution of the beam. The initial and newly coated beam particles in "c" are represented by black and red dots, respectively. The beam particles from the initial beam captured by the mini-bucket are shown by the black region at the center of phase-space distribution. The newly arrived particles and the particles from the initial distribution outside the mini-buckets mix well after coating.

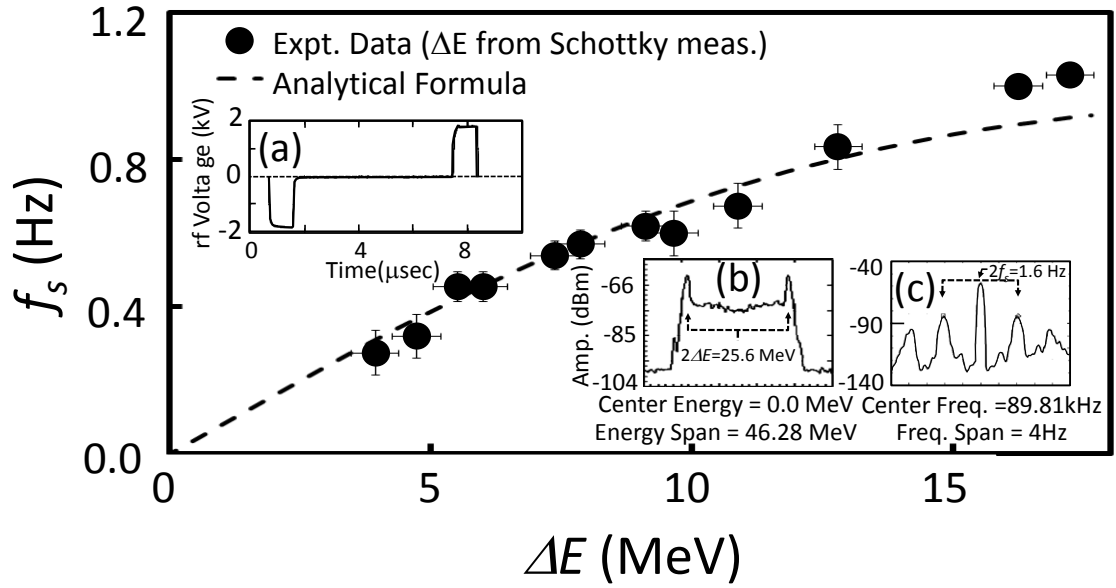

Fig. 4. Measured and calculated incoherent synchrotron frequency spectrum. The insets (a) shows the exact rf waveform and, (b) and (c) show Schottky and Vector Signal Analyzer data for the 12.8 MeV data point, respectively, as illustrations.

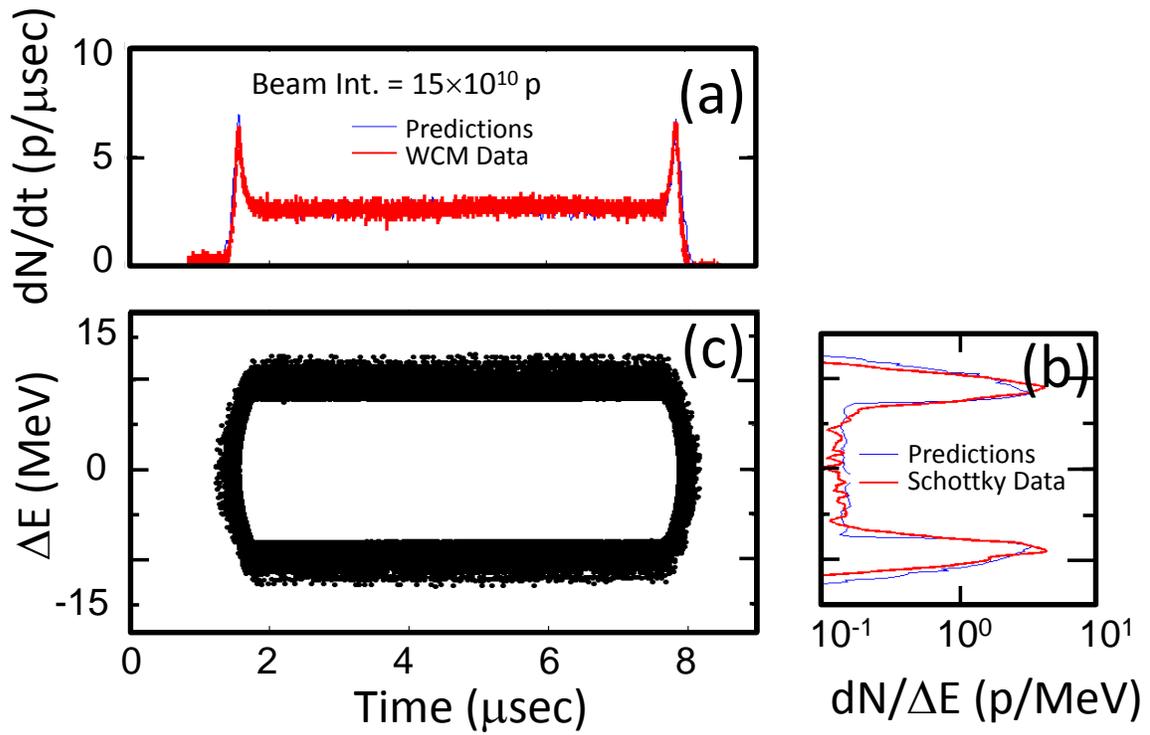

Fig. 5. An overlay of experimental (red curve) and simulation (blue curve) results on hollow beam. The description of the figures is similar to that presented in Fig. 3. The beam particle distribution in "c" is similar to the one shown by red region in Fig. 3(c) but after a single coating.